\documentclass[twocolumn,aps,prd,showpacs,amsmath,amssymb,floatfix,nofootinbib]{revtex4-2}
\usepackage{graphicx}
\usepackage{dcolumn}
\usepackage{bm}
\usepackage{xcolor}
\usepackage{booktabs}
\usepackage{float}
\usepackage{hyperref}

\begin{document}

\title{Non-minimally Coupled Running Curvaton for DESI-motivated Dynamical Dark Energy}

\author{Bichu Li}
\email{libichu@mail.ustc.edu.cn}
\affiliation{Department of Physics, College of Physics, Mechanical and Electrical Engineering,
Jishou University, Jishou 416000, China}

\author{Lei-Hua Liu}
\email{liuleihua8899@hotmail.com, corresponding author}
\affiliation{Department of Physics, College of Physics, Mechanical and Electrical Engineering,
Jishou University, Jishou 416000, China}

\date{\today}

\begin{abstract}
Recent DESI baryon-acoustic-oscillation measurements, when combined with CMB and supernova distance data, have renewed interest in dynamical dark energy whose effective equation of state can cross the phantom divide. We investigate whether such behavior can be embedded in the running-curvaton framework by adding a Jordan-frame non-minimal coupling, $\xi\chi^2R$. The same field is used in two regimes: as a curvaton sector in the early universe and as a late-time dark-energy component. We show that the leading early-time curvaton predictions can be preserved by retuning the effective mass parameter, while the geometric terms in the late-time pressure allow phantom-crossing background histories without introducing a phantom kinetic term. We then perform a background-level likelihood analysis using numerical solutions of the model's late-time homogeneous background equations, a reduced CMB prior on $(\theta_\ast,\omega_b,\omega_{bc})$, the official DESI DR2 BAO data vector with full covariance, and Pantheon+ supernovae without SH0ES calibration. This likelihood tests acoustic-scale, BAO-distance and supernova-distance consistency under the assumption that the curvaton/NMC sector is negligible near recombination. The MCMC analysis gives stable derived background parameters, with $H_0=67.88^{+0.53}_{-0.61}\,{\rm km\,s^{-1}\,Mpc^{-1}}$, $\Omega_m=0.3072^{+0.0059}_{-0.0054}$, $w_0=-0.922^{+0.055}_{-0.063}$ and $w_a=-0.205^{+0.173}_{-0.182}$. The microscopic non-minimal-coupling parameters remain broad, skewed and long-tailed in the marginalized posterior, reflecting degeneracies under background-level data. These results should not be interpreted as a full perturbation-level CMB constraint on the non-minimally coupled curvaton. They are a reduced-CMB+BAO+SN consistency test of the homogeneous expansion history. A Boltzmann-code implementation including TT/TE/EE spectra, CMB lensing, late integrated Sachs-Wolfe effects and scalar perturbation evolution is needed for a definitive test.
\end{abstract}

\maketitle

\section{Introduction}

The standard cosmological model, $\Lambda$CDM, provides a successful description of the CMB anisotropies, baryon-acoustic-oscillation (BAO) distances, supernova luminosity distances and large-scale structure \cite{Planck:2018vyg,SDSS:2003eyl,DES:2021wwk}. The physical origin of the late-time acceleration, however, remains open. Recent DESI DR2 BAO measurements, especially when combined with CMB information and Type-Ia supernova samples such as Pantheon+, Union3 and the DES five-year supernova sample, have sharpened interest in time-dependent dark energy \cite{DESI:2025zgx,DESI:2025wyn,Scolnic:2021amr,Brout:2022vxf,Rubin:2023zlo,DES:2024tys,Wang:2024pmn,Colgain:2024xqj}. In phenomenological fits, this question is often summarized by the Chevallier-Polarski-Linder (CPL) form
\begin{equation}
w(a)=w_0+w_a(1-a),
\end{equation}
introduced in Refs.~\cite{Chevallier:2000qy,Linder:2002et}. The DESI-motivated region can prefer $w_0>-1$ and $w_a<0$, corresponding to an effective equation of state that crosses the phantom divide during the recent expansion history \cite{DESI:2025wyn,DESI:2025zgx,Rubin:2023zlo,Brout:2022vxf,DES:2024tys,Colgain:2024xqj,Pan:2025psn}.

Such a crossing is theoretically nontrivial. Canonical single-field quintessence has a standard kinetic term and is normally restricted to $w\geq -1$ \cite{Caldwell:1997ii,Zlatev:1998tr,Copeland:2006wr}. A field with a wrong-sign kinetic term can realize $w<-1$, but it is associated with ghost instabilities and related pathologies \cite{Caldwell:2003vq,Nojiri:2005sx}. More general single-field constructions can also encounter difficulties when crossing $w=-1$ \cite{Vikman:2004dc,Hu:2004kh}. Stable effective descriptions therefore require extra structure, such as additional degrees of freedom, effective-field-theory operators, modified gravity or scalar-tensor dynamics \cite{Feng:2004ad,Creminelli:2008wc,Gubitosi:2012hu,Wolf:2019hzy,Creminelli:2019kjy,Cai:2025mas,Yang:2024kdo,Yang:2025mws}.

Non-minimal coupling is one such route. Scalar fields coupled to curvature have long been studied in inflation, quintessence and scalar-tensor dark energy \cite{Faraoni:2000wk,Elizalde:2004mq,Gannouji:2006ch,Polarski:2008zp,Perivolaropoulos:2005yv}. In modern language, these theories are related to the broader scalar-tensor/Horndeski class \cite{Horndeski:1974wa,Deffayet:2011gz,Kobayashi:2011nu}. Recent analyses have revisited non-minimal coupling in the context of DESI BAO and dynamical-dark-energy constraints \cite{Ye:2024ywg,Ye:2024thawing,Wolf:2025jed,Pan:2025psn}. Thus the mechanism is not new by itself; the question addressed here is whether it can be incorporated into the running-curvaton framework in a way that is consistent with both its early-time motivation and the late-time distance data.

The running-curvaton model was proposed as a way to connect the early curvaton mechanism with late-time acceleration \cite{Liu:2020running}. In the minimally coupled case, however, the late-time scalar behaves like a thawing canonical field and does not naturally realize the DESI-motivated phantom-crossing behavior \cite{Wolf:2023uno,Wolf:2024eph}. We therefore add a Jordan-frame non-minimal coupling, $\xi\chi^2R$, to the running-curvaton field. The same field then has two roles: during the early universe it sources curvaton perturbations, while at late times it contributes to the dark-energy background through its potential and geometric coupling.

A useful model must satisfy two consistency requirements. First, the non-minimal coupling should not spoil the leading curvaton predictions for the primordial spectrum and non-Gaussianity. We show that the curvature-induced shift of the effective curvaton mass can be absorbed, at leading order, into a retuning of the running-curvaton mass parameter, leaving the standard estimates for $n_s$ and $f_{NL}$ compatible with Planck constraints \cite{Planck:2018jri}. Second, the late-time background should remain compatible with the distance information that motivates the phantom-crossing discussion. We test this point with a reduced CMB prior on $(\theta_\ast,\omega_b,\omega_{bc})$, the official DESI DR2 BAO data vector with its full covariance, and Pantheon+ supernovae without SH0ES/Cepheid calibration \cite{DESI:2025zgx,Scolnic:2021amr,Brout:2022vxf,Torrado:2020dgo}. This is a test of the homogeneous expansion history and the compressed acoustic scale, not of the full CMB anisotropy spectra.

We also state the limitation of the analysis from the outset. The likelihood used below is a background-level reduced-CMB+BAO+SN test. It uses numerical solutions of the model's late-time homogeneous background equations, while the CMB information is compressed to the early-background quantities $(\theta_\ast,\omega_b,\omega_{bc})$. It constrains the homogeneous expansion history and derived quantities such as $H_0$, $\Omega_m$, $r_d$, $w_0$ and $w_a$. It should not be read as a calculation of the full TT/TE/EE spectra, CMB lensing, late ISW effects or scalar perturbation evolution in a non-minimally coupled curvaton theory. The stability discussion is similarly limited: the canonical kinetic structure and luminal tensor speed are useful background-theory checks, but they are not a full perturbation-level parameter-estimation stability analysis. Solar-System and screening constraints require a separate post-Newtonian or environmental analysis \cite{Horndeski:1974wa,Deffayet:2011gz,Kobayashi:2011nu,Will:2014kxa}.


The paper is organized as follows. Section \ref{the model} defines the non-minimally coupled running-curvaton model. Section \ref{early universe consistency} explains how the leading early-universe curvaton predictions can be preserved by parameter re-tuning. Section \ref{late time constraints} derives the late-time background equations, the effective equation of state, and the phantom-crossing condition. Section \ref{sec:reduced_cmb_bao_sn} presents the background-level reduced-CMB+DESI DR2 BAO+Pantheon+ likelihood analysis. Section \ref{discussion} summarizes the results and limitations.

\section{The Model}

\label{the model}

We consider a running-curvaton field $\chi$ non-minimally coupled to gravity.
The Jordan-frame action is
\begin{equation}
\begin{aligned}
S &= \int d^4x \sqrt{-g}
\left[
\frac{1}{2}\left(M_P^2-\xi\chi^2\right)R
-\frac{1}{2}(\nabla\chi)^2
-V(\chi)
\right]\\
&\quad +S_m ,
\end{aligned}
\label{eq:action}
\end{equation}
where $M_P$ is the reduced Planck mass, $\xi$ is the non-minimal coupling and
$S_m$ denotes the matter sector. We adopt the running-curvaton framework of
Ref.~\cite{Liu:2020running}, in which the curvaton mass evolves through an
explicit inflaton-curvaton coupling. The potential is taken to be
\begin{equation}
V(\chi) = \frac{1}{2}\frac{g}{M_P^2}V(\phi)\chi^2+V_1(1-V_0 e^{-\lambda\chi/M_P}),
\label{curvaton potential}
\end{equation}
where $\phi$ is the inflaton. The first term controls the curvaton mass during
the preheating and early-curvaton epochs; it is motivated by the
inflaton-curvaton interaction considered in the running-curvaton scenario and
by parametric-resonance physics during preheating
\cite{Kofman:1994rk,Kofman:1997ms,Liu:2020running}. After the inflaton
decays, this term vanishes. The second term, parameterized by $V_1$, $V_0$ and
$\lambda$, survives at late times and sources the dark-energy-like evolution.

\paragraph{Motivation for the modified potential.}
The choice of the late-time potential in Eq.~\eqref{curvaton potential} is
phenomenological. In the original minimally coupled running-curvaton setup,
the pure exponential potential leads to late-time dynamics similar to thawing
quintessence \cite{Wolf:2023uno}, whose trajectories are difficult to align
with the DESI-motivated CPL region \cite{DESI:2025wyn,DESI:2025zgx}. By
contrast, the modified form
\[
V(\chi)=V_1\left(1-V_0 e^{-\lambda \chi/M_P}\right)
\]
introduces additional shape freedom that allows the field evolution to
interpolate between a nearly frozen stage and a later rolling phase. When
combined with $\xi\chi^2R$, this structure more readily generates
phantom-crossing trajectories in the DESI-motivated region with $w_0>-1$ and
$w_a<0$ \cite{Wolf:2024eph}. The potential should therefore be understood as a
phenomenological extension used to test whether a running-curvaton model can
realize such late-time dynamics while preserving its early-time motivation.

The physical picture is then as follows. During preheating, part of the
inflaton energy can be transferred to the curvaton through the coupling in
Eq.~\eqref{curvaton potential}; the curvaton subsequently sources the
primordial curvature perturbation. Because it is long-lived, the same field
can remain relevant at late times, where the second term in
Eq.~\eqref{curvaton potential} dominates its potential energy. The
non-minimal coupling then allows the effective dark-energy equation of state
to cross $w=-1$ at the background level.

\section{Early-time consistency of the curvaton sector}
\label{early universe consistency}
A non-minimal coupling changes the effective curvaton mass through the Ricci
scalar. It is therefore important to specify under which assumptions the
early-time predictions of the running-curvaton mechanism remain intact. We
use the spatially flat Friedmann-Lemaitre-Robertson-Walker metric
\begin{equation}
ds^2 = -dt^2 + a^2(t) \delta_{ij} dx^i dx^j ,
\label{frw metric}
\end{equation}
where $a(t)$ is the scale factor. The discussion in this section is a
leading slow-roll consistency argument. It does not compute the full
perturbation-level CMB likelihood for the non-minimally coupled theory.

\subsection{Effective mass and retuning}

Expanding the Jordan-frame action in Eq.~\eqref{eq:action} to quadratic
order in $\delta\chi$ shows that the mass term of the curvaton fluctuation
is shifted by the curvature coupling:
\begin{equation}
m_{\rm eff}^2 = m_{\rm orig}^2 + \xi R .
\label{effective mass of curvaton}
\end{equation}
For the metric in Eq.~\eqref{frw metric},
\begin{equation}
R = 6(2H^2 + \dot{H}) \,,
\label{ricci scalar}
\end{equation}
and during slow-roll inflation $R\simeq 12H^2$. The original running-curvaton
mass is determined by the inflaton-curvaton coupling and may be written as
$m_{\rm orig}^2\simeq 6g_0H^2$ in the notation of
Ref.~\cite{Liu:2020running}. Hence
\begin{equation}
m_{\rm eff}^2 = 6g_0 H^2 + 12\xi H^2 = 6H^2 (g_0 + 2\xi),
\label{effective mass of curvaton1}
\end{equation} 
The observable curvaton tilt therefore depends on the effective combination
$g_0+2\xi$. We define
\begin{equation}
g_0^{\rm obs}\equiv g_0+2\xi ,
\label{eq:g0obs}
\end{equation}
and retune $g_0$ so that $g_0^{\rm obs}$ matches the value required in the
minimally coupled running-curvaton model. The de-Sitter estimate of the
curvaton spectral index then becomes
\begin{equation}
n_{\chi}-1\simeq 4(g_0+2\xi)=4g_{0}^{\rm obs}.
    \label{spectral index of curvaton final}
\end{equation}
The detailed spectral-index derivation is given in
Appendix~\ref{app:spectral_index}. This retuning argument is a leading-order
early-time consistency condition. It does not replace a full perturbative
CMB likelihood.

\subsection{Non-Gaussianity and tensor modes}

The local non-Gaussianity parameter $f_{\rm NL}$ in curvaton models is
controlled mainly by the curvaton energy fraction at decay,
$r_{\rm dec}\equiv 3\rho_\chi/(3\rho_\chi+4\rho_\gamma)$
\cite{Liu:2020running}. In the sudden-decay approximation, reviewed in
Appendix~\ref{sec:fnl_calculation}, one obtains
\begin{equation}
f_{\rm NL} \approx \frac{5}{4r_{\rm dec}} \,,
\end{equation}
which is compatible with Planck bounds for sufficiently large
$r_{\rm dec}$ \cite{Planck:2018jri}. The retuning in Eq.~\eqref{eq:g0obs}
keeps the inflationary curvaton fluctuation sector fixed at leading order,
and during radiation domination $R\simeq0$, so the explicit curvature-induced
mass correction is suppressed at decay.

Tensor perturbations are discussed in Appendix~\ref{tensor modes}. The
non-minimal coupling changes the effective Planck mass by
$\Delta M_{\rm eff}^2=-\xi\chi^2$. Because the curvaton is subdominant during
inflation, the correction to the tensor amplitude is of order
$\xi\chi^2/M_P^2$ and the usual inflationary tensor prediction is recovered
to leading order. The central limitation remains that these arguments address
the leading early-time curvaton observables, while the full CMB anisotropy
spectra require a Boltzmann-code treatment of the coupled background and
perturbations.

\section{Late-time background dynamics}
\label{late time constraints}

After the inflaton has decayed, the first term in
Eq.~\eqref{curvaton potential} is negligible and the curvaton evolves as a
late-time scalar-tensor degree of freedom. The same non-minimal coupling that
shifts the early curvaton mass then modifies the effective dark-energy
pressure. This section collects the background equations and the theoretical
conditions used before the model is confronted with data.

\subsection{Modified Friedmann equations}

Varying the action in Eq.~\eqref{eq:action} on a flat FLRW background gives
\begin{equation}
\begin{aligned}
3(M_P^2-\xi\chi^2)H^2
&=\rho_m+\frac{1}{2}\dot{\chi}^2+V(\chi)
 +6\xi H\chi\dot{\chi},
\end{aligned}
\label{eq:friedmann_late_1}
\end{equation}
and
\begin{align}
-2(M_P^2-\xi\chi^2)\dot H
&=\rho_m+\dot{\chi}^2+2\xi H\chi\dot{\chi} \nonumber\\
&\quad -2\xi(\dot{\chi}^2+\chi\ddot{\chi}) .
\label{eq:friedmann_late_2}
\end{align}
The scalar-field equation is
\begin{equation}
\ddot{\chi}+3H\dot{\chi}+V_{,\chi}+\xi R\chi=0 .
\label{eq:late_scalar_eom}
\end{equation}
The effective Planck mass,
\begin{equation}
M_{\rm eff}^2=M_P^2-\xi\chi^2 ,
\end{equation}
must remain positive along the numerical trajectory.

\subsection{Effective equation of state and CPL projection}

For comparison with dark-energy phenomenology, we rewrite the background
equations in the standard Einstein form
$3M_P^2H^2=\rho_m+\rho_{\rm DE}^{\rm eff}$. The effective dark-energy density
and pressure are
\begin{equation}
\rho_{\rm DE}^{\rm eff}
=\frac{1}{2}\dot{\chi}^2+V(\chi)
+6\xi H\chi\dot{\chi}+3\xi\chi^2H^2 ,
\label{energy density of de}
\end{equation}
and
\begin{align}
p_{\rm DE}^{\rm eff}
&=\frac{1}{2}\dot{\chi}^2-V(\chi)
-2\xi(\chi\ddot{\chi}+\dot{\chi}^2) \nonumber\\
&\quad -4\xi H\chi\dot{\chi}
-\xi\chi^2(2\dot H+3H^2).
\label{presure of de}
\end{align}
We define
\begin{equation}
w_{\rm eff}=\frac{p_{\rm DE}^{\rm eff}}{\rho_{\rm DE}^{\rm eff}} .
\end{equation}
In the redshift variable, $1+z=1/a$ and
$\dot{\chi}=-(1+z)H\chi'$, where a prime denotes $d/dz$. The numerical
background evolution can be written as
\begin{equation}
H^2(z)=\frac{\rho_{m,0}(1+z)^3+V(\chi)}{{\cal D}(z)},
\label{Hubble parameter in terms of z}
\end{equation}
where
\begin{align}
{\cal D}(z)
&=3(M_P^2-\xi\chi^2)-\frac{1}{2}(1+z)^2\chi'^2 \nonumber\\
&\quad +6\xi(1+z)\chi\chi' .
\end{align}
The derivative of $H$ is evaluated as
\begin{equation}
H'(z)=\frac{S_1(z)-2\xi\chi S_2(z)}{(1+z)H\Delta},
\label{derivative of H for z}
\end{equation}
with
\begin{align}
S_1(z)
&=\rho_{m0}(1+z)^3
 +(1-2\xi)(1+z)^2H^2\chi'^2 \nonumber\\
&\quad -2\xi(1+z)H^2\chi\chi',\\
S_2(z)
&=3(1+z)H^2\chi'-V_{,\chi}-12\xi H^2\chi,\\
\Delta
&=2(M_P^2-\xi\chi^2+6\xi^2\chi^2).
\end{align}
The CPL parameters quoted later are extracted from the 
numerically solved late-time background through
\begin{equation}
w_0=w_{\rm eff}(z=0),\qquad
w_a=\left.\frac{d w_{\rm eff}}{dz}\right|_{z=0},
\label{eq:w0wa_def}
\end{equation}
which matches the usual convention $w(a)=w_0+w_a(1-a)$ at $z=0$.

\subsection{Phantom crossing and theoretical viability}

For a canonical minimally coupled scalar, $w_{\rm eff}\geq -1$. In the present
model the crossing condition is modified because the effective pressure
contains geometric terms. Combining Eqs.~\eqref{energy density of de} and
\eqref{presure of de}, one obtains
\begin{equation}
\begin{aligned}
\rho_{\rm DE}^{\rm eff}+p_{\rm DE}^{\rm eff}
&=(1-2\xi)\dot{\chi}^2\\
&\quad +2\xi
\left(H\chi\dot{\chi}-\chi\ddot{\chi}-\chi^2\dot H\right).
\end{aligned}
\label{eq:phantom_crossing_condition}
\end{equation}
The second line can become sufficiently negative during part of the
late-time evolution, allowing $w_{\rm eff}$ to cross $-1$ without changing
the sign of the fundamental kinetic term.

The action in Eq.~\eqref{eq:action} belongs to the scalar-tensor/Horndeski
class \cite{Horndeski:1974wa}, with
\begin{align}
G_2(\chi,X)&=X-V(\chi),\\
G_4(\chi,X)&=\frac{1}{2}(M_P^2-\xi\chi^2),\\
G_3(\chi,X)&=G_5(\chi,X)=0,
\end{align}
where $X\equiv-\nabla_\mu\chi\nabla^\mu\chi/2$. Since $G_{4,X}=0$, the
tensor propagation speed is luminal, $c_T^2=1$. The canonical kinetic
structure and $G_{4,X}=0$ indicate that no explicit phantom kinetic term is
introduced at the background-theory level. However, a full perturbation-level
stability analysis would require checking the scalar kinetic and gradient
matrices along the sampled trajectories. In the numerical analysis we reject
parameter points for which $M_{\rm eff}^2\leq0$ or the Friedmann denominator
becomes non-positive.

Local gravity constraints require a separate analysis. In high-density
environments the scalar mass may receive a contribution of order
\begin{equation}
m_{\rm loc}^2\simeq V''(\chi)+\xi\frac{\rho_{\rm loc}}{M_P^2},
\end{equation}
suggesting possible Yukawa suppression of fifth-force effects, in analogy
with screening scenarios such as chameleon or symmetron screening
\cite{Khoury:2003aq,Hinterbichler:2010es,Khoury:2010xi}. This statement is
only qualitative; a static solution and a post-Newtonian analysis are needed
for a definitive Solar-System test. The cosmological value of
$F(0)=M_P^2-\xi\chi_0^2$ should not be directly identified with a locally
measured Newton constant without specifying the local scalar profile and
possible environmental screening. We therefore do not claim Solar-System
viability or an explicit screening mechanism in the present work.

\section{Background-level reduced-CMB+BAO+SN consistency test}
\label{sec:reduced_cmb_bao_sn}

\subsection{Purpose and limitations}

The previous section specifies the homogeneous background evolution and the
mapping to the derived quantities $w_0$ and $w_a$. We now test that
background against distance information from a compressed CMB prior, the
official DESI DR2 BAO data vector and the Pantheon+ supernova sample. The
late-time evolution entering the likelihood is obtained from
numerical solutions of the model's homogeneous background equations. The
CMB part, however, is reduced to the
early-background/acoustic-scale prior on
$(\theta_\ast,\omega_b,\omega_{bc})$. This analysis therefore answers a
limited question: under the reduced-CMB assumptions, can the same model that
preserves the leading curvaton predictions also give a viable late-time
expansion history? It does not answer the separate question of whether the
model fits the full CMB anisotropy spectra.

We emphasize this limitation throughout the paper. The calculation below
uses the numerical late-time homogeneous background solution of
the model, but the CMB part is compressed to the early-Universe quantities
$(\theta_\ast,\omega_b,\omega_{bc})$. It can check compatibility with the
acoustic scale, BAO distances and SN luminosity distances if the
curvaton/NMC sector is negligible around recombination. It does not include
the effects of the non-minimal coupling on the full temperature and
polarization spectra, CMB lensing, the late integrated Sachs-Wolfe effect,
or scalar perturbation evolution. Those effects require a Boltzmann solver
and are left for future work.

\subsection{Compressed CMB prior}

Following Appendix A of the DESI DR2 cosmology analysis
\cite{DESI:2025zgx}, and the early-Universe CMB compression of
Ref.~\cite{Lemos:2023xhs}, we impose a Gaussian prior on
\begin{equation}
\mathbf{x}_{\rm CMB}
=
\left(
\theta_\ast,\omega_b,\omega_{bc}
\right),
\end{equation}
where
\begin{equation}
\omega_b=\Omega_b h^2,\qquad
\omega_c=\Omega_c h^2,\qquad
\omega_{bc}=\omega_b+\omega_c .
\end{equation}
The acoustic angle is computed from the numerical background as
\begin{equation}
\theta_\ast=\frac{r_s(z_\ast)}{D_M(z_\ast)} ,
\end{equation}
where $D_M$ is the transverse comoving distance. The sound horizon entering
the acoustic angle is evaluated as
\begin{equation}
r_s(z)=\frac{c}{H_0}\int_z^\infty
\frac{c_s(z')}{E_{\rm std}(z')}\,dz',
\qquad
c_s=\frac{1}{\sqrt{3(1+R_b)}} ,
\end{equation}
with
\begin{equation}
R_b(z)=\frac{3\Omega_b}{4\Omega_\gamma(1+z)} .
\end{equation}
The recombination redshift $z_\ast$ is evaluated with the Hu-Sugiyama
fitting formula \cite{Hu:1995en}, while the drag redshift $z_d$ is evaluated
with the Eisenstein-Hu fitting formula \cite{Eisenstein:1997ik}. In the
sound-horizon integral we use the standard high-redshift radiation-plus-matter
background. The fixed early-universe constants are
$T_{\rm CMB}=2.7255\,{\rm K}$ and $N_{\rm eff}=3.046$, with
\begin{equation}
\begin{aligned}
\omega_\gamma
&=2.469\times10^{-5}
\left(\frac{T_{\rm CMB}}{2.7255\,{\rm K}}\right)^4,\\
\omega_r
&=\omega_\gamma(1+0.227107\,N_{\rm eff}) .
\end{aligned}
\end{equation}
No independent neutrino-mass parameter is varied in this reduced-background
analysis. The BAO sound horizon used in the likelihood is the same
DESI-style fitting formula,
\begin{equation}
\begin{aligned}
r_d&=147.05\,{\rm Mpc}
\left(\frac{\omega_b}{0.02236}\right)^{-0.13}\\
&\quad\times
\left(\frac{\omega_{bc}}{0.1432}\right)^{-0.23}
\left(\frac{N_{\rm eff}}{3.04}\right)^{-0.10}.
\end{aligned}
\end{equation}
Thus the implementation is a hybrid reduced-background prescription:
$r_s(z_\ast)$ is evaluated in the standard high-redshift
radiation-plus-matter background, $r_d$ is given by the DESI-style fitting
formula, and the late-time part of the distance $D_M(z_\ast)$ is controlled
by the model background while the high-redshift part follows the same
standard early-background assumption. These choices make the CMB prior an
acoustic-scale and early-background calibration check. They do not compute
recombination physics or the full CMB spectra.
The mean and covariance are
\begin{equation}
\boldsymbol{\mu}_{\rm CMB}
=
\begin{pmatrix}
0.01041\\
0.02223\\
0.14208
\end{pmatrix},
\end{equation}
and
\begin{equation}
\mathbf{C}_{\rm CMB}
=
10^{-9}
\begin{pmatrix}
0.006621 & 0.12444 & -1.1929\\
0.12444 & 21.344 & -94.001\\
-1.1929 & -94.001 & 1488.4
\end{pmatrix}.
\end{equation}
The CMB contribution is therefore
\begin{equation}
\chi^2_{\rm CMB}
=
\left(\mathbf{x}_{\rm CMB}-\boldsymbol{\mu}_{\rm CMB}\right)^T
\mathbf{C}_{\rm CMB}^{-1}
\left(\mathbf{x}_{\rm CMB}-\boldsymbol{\mu}_{\rm CMB}\right).
\end{equation}
The third entry is the total physical baryon-plus-cold-dark-matter density
$\omega_{bc}$, not $\omega_c$ alone.

\subsection{DESI DR2 BAO and Pantheon+ supernova likelihoods}

For BAO we use the public DESI DR2 all-tracer Gaussian BAO data vector and
its full covariance matrix. The files are read from the Cobaya/CobayaSampler
data package \cite{Torrado:2020dgo}, which is used here as a data and
likelihood packaging source rather than as a separate observational data set;
the measurements themselves are the DESI DR2 BAO measurements described in
Ref.~\cite{DESI:2025zgx}. The measurement and covariance files are
{\tt desi\_gaussian\_bao\_ALL\_GCcomb\_mean.txt} and
{\tt desi\_gaussian\_bao\_ALL\_GCcomb\_cov.txt}, respectively. The BAO data
vector contains 13 entries and the
covariance matrix is $13\times 13$, symmetric and positive definite. The
model predictions are evaluated with the same sound horizon prescription
used in the CMB and BAO parts of the likelihood. The supported BAO
quantities are
\begin{equation}
\begin{gathered}
D_M/r_d,\qquad D_H/r_d,\qquad D_V/r_d,\\
D_A/r_d,\qquad H(z)r_d ,
\end{gathered}
\end{equation}
where $H(z)$ is in ${\rm km\,s^{-1}\,Mpc^{-1}}$ and $r_d$ is in Mpc.
Thus the BAO contribution is
\begin{equation}
\chi^2_{\rm BAO}
=
\Delta_{\rm BAO}^{T}\mathbf{C}_{\rm BAO}^{-1}\Delta_{\rm BAO}.
\end{equation}
The data-vector entries used in this likelihood are listed in
Table~\ref{tab:bao_dr2_vector}. The present DESI DR2 vector contains
$D_V/r_d$, $D_M/r_d$ and $D_H/r_d$ entries. The code also supports
$D_A/r_d$ and $H(z)r_d$ for other BAO data sets, with $H(z)r_d$ interpreted
in ${\rm km\,s^{-1}}$.

\begin{table}[H]
\centering
\scriptsize
\setlength{\tabcolsep}{4pt}
\caption{DESI DR2 all-tracer BAO data vector used in the baseline likelihood, read from {\tt desi\_gaussian\_bao\_ALL\_GCcomb\_mean.txt}. The covariance is the corresponding full $13\times13$ matrix in {\tt desi\_gaussian\_bao\_ALL\_GCcomb\_cov.txt}.}
\label{tab:bao_dr2_vector}
\begin{tabular}{ccc}
\toprule
$z_{\rm eff}$ & Observable & Value\\
\midrule
0.295 & $D_V/r_d$ & 7.9417\\
0.510 & $D_M/r_d$ & 13.5876\\
0.510 & $D_H/r_d$ & 21.8629\\
0.706 & $D_M/r_d$ & 17.3507\\
0.706 & $D_H/r_d$ & 19.4553\\
0.934 & $D_M/r_d$ & 21.5756\\
0.934 & $D_H/r_d$ & 17.6415\\
1.321 & $D_M/r_d$ & 27.6009\\
1.321 & $D_H/r_d$ & 14.1760\\
1.484 & $D_M/r_d$ & 30.5119\\
1.484 & $D_H/r_d$ & 12.8170\\
2.330 & $D_H/r_d$ & 8.6315\\
2.330 & $D_M/r_d$ & 38.9890\\
\bottomrule
\end{tabular}
\end{table}

For SNe we use the Pantheon+ public data and covariance release, described
in the Pantheon+ data-release paper \cite{Scolnic:2021amr}, together with
the cosmological-likelihood treatment of the Pantheon+ analysis
\cite{Brout:2022vxf}. The files are read through the corresponding public
data package. As for the BAO case, the package is a convenient source for
the public data vector and covariance matrix, not a new supernova survey. The
data and covariance files are
{\tt Pantheon+SH0ES.dat} and
{\tt Pantheon+SH0ES\_STAT+SYS.cov}. In the baseline run we use Pantheon+
without SH0ES: we keep objects with finite $z_{\rm HD}$ and corrected
magnitude, impose $z_{\rm HD}\geq 0.01$, and remove Cepheid calibrator rows
by requiring ${\tt IS\_CALIBRATOR}\leq0$. This leaves 1580 objects out of
the original 1701 rows; the covariance matrix is reduced with the same
Boolean mask and in the same row order. No SH0ES Cepheid calibration prior
is included in the baseline analysis. Pantheon+ is adopted as the baseline
supernova sample in this work. A full comparison with Union3 and the DES
five-year supernova sample is left for future robustness analyses.

For the SN redshift convention, the cosmological redshift $z_{\rm HD}$ is used
in the comoving-distance integral, while the heliocentric redshift
$z_{\rm HEL}$ enters the photon-energy factor in the luminosity distance. Thus
the theoretical distance modulus is
\begin{equation}
\mu_{\rm th}
=5\log_{10}\left[
\frac{(1+z_{\rm HEL})D_M(z_{\rm HD})}{\rm Mpc}
\right]+25+\Delta_M ,
\end{equation}
where $\Delta_M$ is a nuisance magnitude offset. We minimize analytically
over this offset. If
\begin{equation}
\delta=\mu_{\rm model}-\mu_{\rm obs},\qquad
\mathbf{1}=(1,\ldots,1)^T ,
\end{equation}
then
\begin{equation}
\begin{aligned}
A&=\delta^T\mathbf{C}_{\rm SN}^{-1}\delta,\\
B&=\mathbf{1}^T\mathbf{C}_{\rm SN}^{-1}\delta,\\
C&=\mathbf{1}^T\mathbf{C}_{\rm SN}^{-1}\mathbf{1},
\end{aligned}
\end{equation}
and
\begin{equation}
\chi^2_{\rm SN}=A-\frac{B^2}{C},\qquad
\Delta_M^{\rm best}=-\frac{B}{C}.
\end{equation}
Here $A$ is the chi-square before fitting the common magnitude offset, $B$
measures the projection of the residual vector onto a constant magnitude
shift, and $C$ is the inverse-variance weight of that constant mode. The
expression above is the result of minimizing the quadratic form with respect
to $\Delta_M$.
The total likelihood used in this section is
\begin{equation}
\begin{aligned}
\chi^2_{\rm tot}
&=\chi^2_{\rm CMB}+\chi^2_{\rm BAO}+\chi^2_{\rm SN},\\
\ln {\cal L}&=-\frac{1}{2}\chi^2_{\rm tot}.
\end{aligned}
\end{equation}

\subsection{Sampled parameters, priors and numerical setup}

The sampled parameters are
\begin{equation}
\Theta=
\left\{
h,\omega_b,\omega_c,\xi,\lambda,V_0,\chi_i
\right\}.
\end{equation}
The late-time normalization $V_1$ is not sampled independently; for every
sample it is calibrated by a shooting condition requiring $E(0)=H(0)/H_0=1$.
The baseline top-hat priors are
\begin{align}
0.5&<h<0.9, &
0.020&<\omega_b<0.0248, \nonumber\\
0.080&<\omega_c<0.160, &
0&<\xi<5, \\
0.05&<\lambda<3, &
0.05&<V_0<2, \nonumber\\
10^{-4}&<\chi_i<0.55. &&
\end{align}
Parameter points are rejected if the effective Planck mass
$F(\chi)=M_P^2-\xi\chi^2$ becomes non-positive, if the Friedmann
denominator becomes non-positive, or if the numerical shooting fails.

We sample the posterior with the affine-invariant ensemble sampler
\texttt{emcee} \cite{Foreman-Mackey:2012any}. The main CMB+BAO+SN chain
uses 56 walkers and 30000 steps. We discard the first 6000 steps as burn-in
and retain every tenth sample for the quoted posterior summaries. The
convergence diagnostics are computed from the full MCMC chain.

\subsection{CMB+BAO+SN posterior}

The best-fit point of the 30000-step run gives
\begin{equation}
\begin{aligned}
\chi^2_{\rm CMB}&=0.065, &
\chi^2_{\rm BAO}&=7.899,\\
\chi^2_{\rm SN}&=1386.133.&
\end{aligned}
\end{equation}
and hence
\begin{equation}
\chi^2_{\rm tot}=1394.098,\qquad
\Delta_M^{\rm best}=-19.424 .
\end{equation}
To assess the fit relative to a reference cosmology, we also fit a flat
$\Lambda$CDM model with the same likelihood pipeline. The reference run uses
the same reduced CMB prior, DESI DR2 BAO data vector and covariance, and
Pantheon+ without SH0ES/Cepheid calibration, with the same analytic
minimization over $\Delta_M$. The fitted $\Lambda$CDM parameters are
$(h,\omega_b,\omega_c)$. Following the usual Gaussian-likelihood convention,
we compare the minimum chi-square values and the Akaike information criterion
\cite{Akaike:1974ic}
\begin{equation}
{\rm AIC}=\chi^2_{\rm min}+2k ,
\end{equation}
where $k$ is the number of fitted cosmological/model parameters. The SN
offset is treated in the same way for both models, so including or excluding
it from $k$ would not change $\Delta{\rm AIC}$. The comparison is summarized
in Table~\ref{tab:model_comparison}. The non-minimally coupled running
curvaton improves the best-fit chi-square by
$\Delta\chi^2=\chi^2_{\rm NMC}-\chi^2_{\Lambda{\rm CDM}}=-6.56$, while the
AIC difference is $\Delta{\rm AIC}=+1.44$ because of the four additional
microscopic parameters. We therefore interpret the result as evidence that
the model is compatible with the distance data and can modestly improve the
best-fit chi-square within this background-level test, but not as a decisive
statistical preference over $\Lambda$CDM by AIC.
This conservative interpretation is in line with recent analyses of
late-time dynamical dark energy and pre-recombination physics
\cite{Gonzalez-Fuentes:2026npx}: phantom-crossing dynamics can improve the
maximum likelihood relative to $\Lambda$CDM, but the statistical reading
depends on parameter penalties, data choices, and assumptions about the
high-redshift sector.

The corresponding best-fit background quantities are
\begin{equation}
\begin{aligned}
H_0&=67.49~{\rm km\,s^{-1}\,Mpc^{-1}},&
\Omega_m&=0.311,\\
\theta_\ast&=0.0104102,&
r_d&=147.46~{\rm Mpc},\\
w_0&=-0.858,&
w_a&=-0.433.
\end{aligned}
\end{equation}
The posterior medians and 68\% credible intervals are summarized in
Table~\ref{tab:cmb_bao_sn_summary}. We quote medians rather than means as the
default summary statistics because several microscopic parameters have
skewed or long-tailed marginalized distributions. The best-fit point is
therefore reported separately through the chi-square values above.

\begin{table}[t]
\centering
\scriptsize
\setlength{\tabcolsep}{3.2pt}
\caption{Posterior medians and 68\% credible intervals from the
background-level reduced-CMB+DESI DR2 BAO+Pantheon+ analysis. Pantheon+ is
used without SH0ES/Cepheid calibration; $H_0$ and $r_d$ are in
${\rm km\,s^{-1}\,Mpc^{-1}}$ and Mpc, respectively.}
\label{tab:cmb_bao_sn_summary}
\begin{tabular}{lccc}
\toprule
Quantity & Median & $-1\sigma$ & $+1\sigma$\\
\midrule
$h$ & 0.67851 & 0.00562 & 0.00552\\
$\omega_b$ & 0.022277 & 0.000124 & 0.000123\\
$\omega_c$ & 0.119201 & 0.000760 & 0.000812\\
$\xi$ & 2.829 & 1.468 & 1.582\\
$\lambda$ & 1.430 & 0.849 & 0.882\\
$V_0$ & 0.497 & 0.225 & 0.199\\
$\chi_i$ & 0.0673 & 0.0455 & 0.0567\\
\midrule
$H_0$ & 67.88 & 0.61 & 0.53\\
$\Omega_m$ & 0.3072 & 0.0054 & 0.0059\\
$\omega_{bc}$ & 0.14146 & 0.00070 & 0.00079\\
$r_d$ & 147.51 & 0.19 & 0.18\\
$\theta_\ast$ & 0.0104107 & 0.0000023 & 0.0000023\\
$w_0$ & -0.922 & 0.063 & 0.055\\
$w_a$ & -0.205 & 0.182 & 0.173\\
$\Delta_M$ & -19.418 & 0.013 & 0.011\\
\bottomrule
\end{tabular}
\end{table}

\begin{table*}[t]
\centering
\scriptsize
\setlength{\tabcolsep}{5.0pt}
\caption{Best-fit chi-square and AIC comparison with a flat $\Lambda$CDM
reference fit using the same reduced-CMB+DESI DR2 BAO+Pantheon+ likelihood.
Here $\Delta\chi^2$ and $\Delta{\rm AIC}$ are quoted relative to the
$\Lambda$CDM row. Pantheon+ is used without SH0ES/Cepheid calibration in both
fits.}
\label{tab:model_comparison}
\begin{tabular}{lcccccccc}
\toprule
Model & $k$ & $\chi^2_{\rm CMB}$ & $\chi^2_{\rm BAO}$ &
$\chi^2_{\rm SN}$ & $\chi^2_{\rm tot}$ &
$\Delta\chi^2$ & AIC & $\Delta{\rm AIC}$\\
\midrule
$\Lambda$CDM & 3 & 0.319 & 10.907 & 1389.433 & 1400.660 & 0 & 1406.660 & 0\\
NMC running curvaton & 7 & 0.065 & 7.899 & 1386.133 & 1394.098 & -6.562 & 1408.098 & 1.438\\
\bottomrule
\end{tabular}
\end{table*}

\subsection{Early-background consistency diagnostics}

The detailed early-background diagnostics are now collected in
Appendix~\ref{app:reduced_cmb_diagnostics}. The main points are as follows.
The best-fit CMB pull vector is small in each diagonal direction of the compressed prior, and the posterior median of $\theta_\ast$ remains centered on the reduced-CMB prior. However, these diagnostics should not be interpreted as a full recombination-era or perturbation-level CMB test. Since the present late-time background solver starts at $z_{\rm ini}=50$, the
reduced-CMB likelihood is an acoustic-scale consistency check under the
assumption that the curvaton/NMC sector is negligible near recombination,
rather than a direct evolution test of $\chi(z_\ast)$, $F(z_\ast)/M_P^2$ or
$\Omega_\chi(z_\ast)$. At the level of the present background approximation,
this assumption is supported by the near-static initial condition for the
late-time field, the large Hubble friction at $z>50$, and the small inferred
fractional contribution at the start of the numerical solve,
$\Omega_\chi(z_{\rm ini}=50)\simeq 0.003$ at the best-fit point and
$\simeq0.008$ for the posterior median. The appendix also reports the
available low-redshift quantities, such as $F(0)/M_P^2$ and the minimum of
$F/M_P^2$ over $0<z<50$. These quantities document the assumptions of the
reduced-CMB consistency check and are not used as additional likelihood
terms. The cosmological value of $F(0)$ should also not be directly
identified with a locally measured Newton constant without specifying the
local scalar profile and possible environmental screening.

\begin{figure*}[t]
\centering
\includegraphics[width=0.44\linewidth]{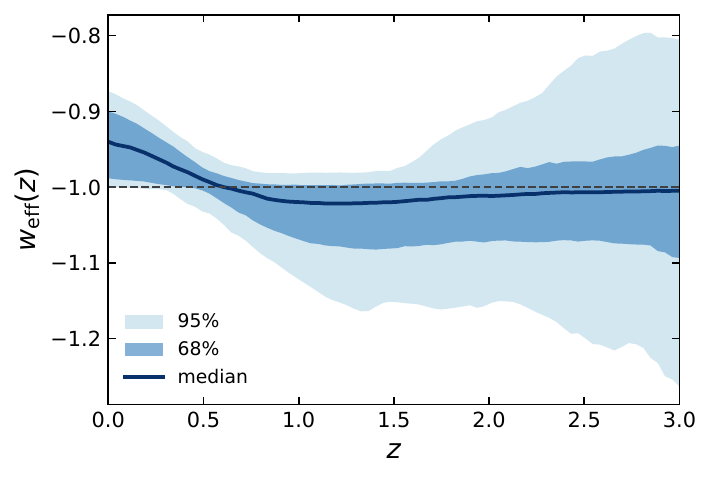}
\includegraphics[width=0.44\linewidth]{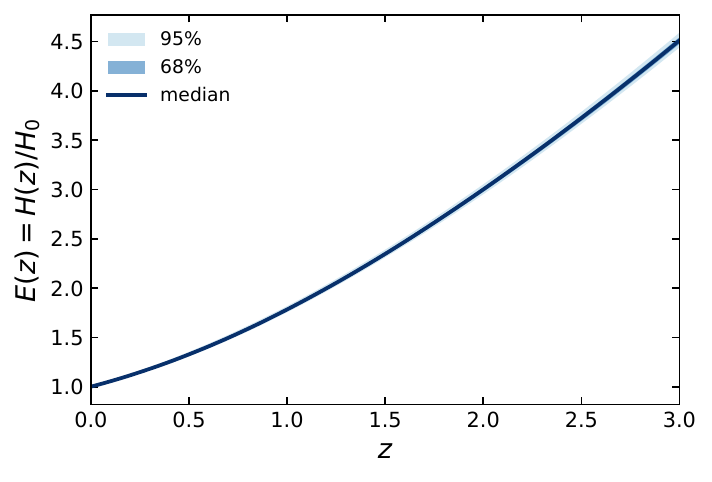}
\caption{Posterior bands for the late-time background evolution inferred
from the reduced-CMB+DESI DR2 BAO+Pantheon+ analysis. Left: effective
equation of state $w_{\rm eff}(z)$, with the dashed line indicating
$w=-1$. Right: normalized expansion rate $E(z)=H(z)/H_0$. The bands enclose
68\% and 95\% of the sampled posterior.}
\label{fig:posterior_bands}
\end{figure*}

\begin{figure*}[t]
\centering
\includegraphics[width=0.82\textwidth]{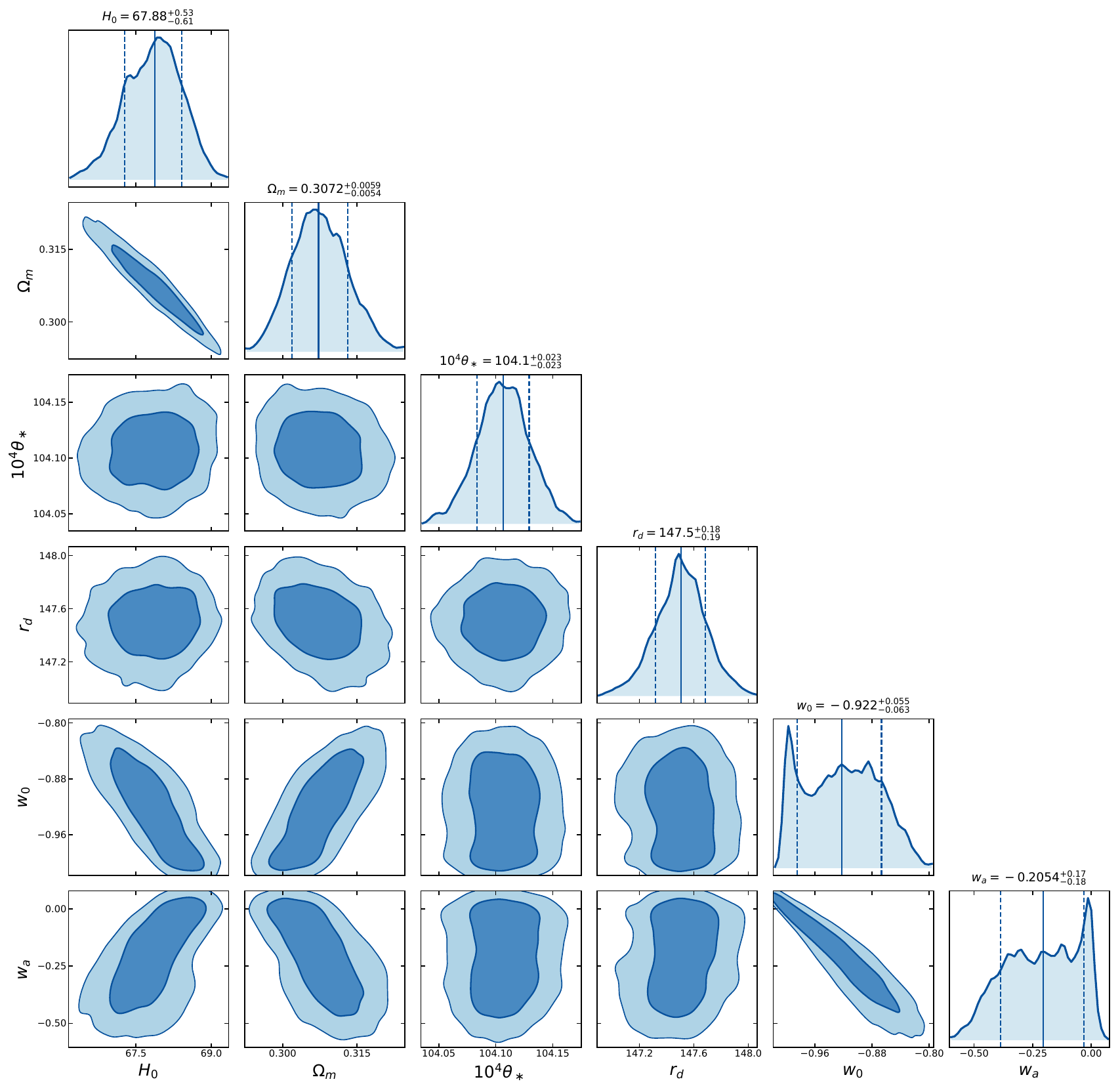}
\caption{Posterior distribution of the main derived background quantities
$(H_0,\Omega_m,\theta_\ast,r_d,w_0,w_a)$ from the reduced-CMB+DESI DR2 BAO+Pantheon+
analysis. The contours show the 68\% and 95\% credible regions. The combined
distance data constrain these background quantities much more tightly than
the microscopic non-minimal-coupling parameters.}
\label{fig:background_posterior}
\end{figure*}

\begin{figure*}[t]
\centering
\includegraphics[width=0.56\textwidth]{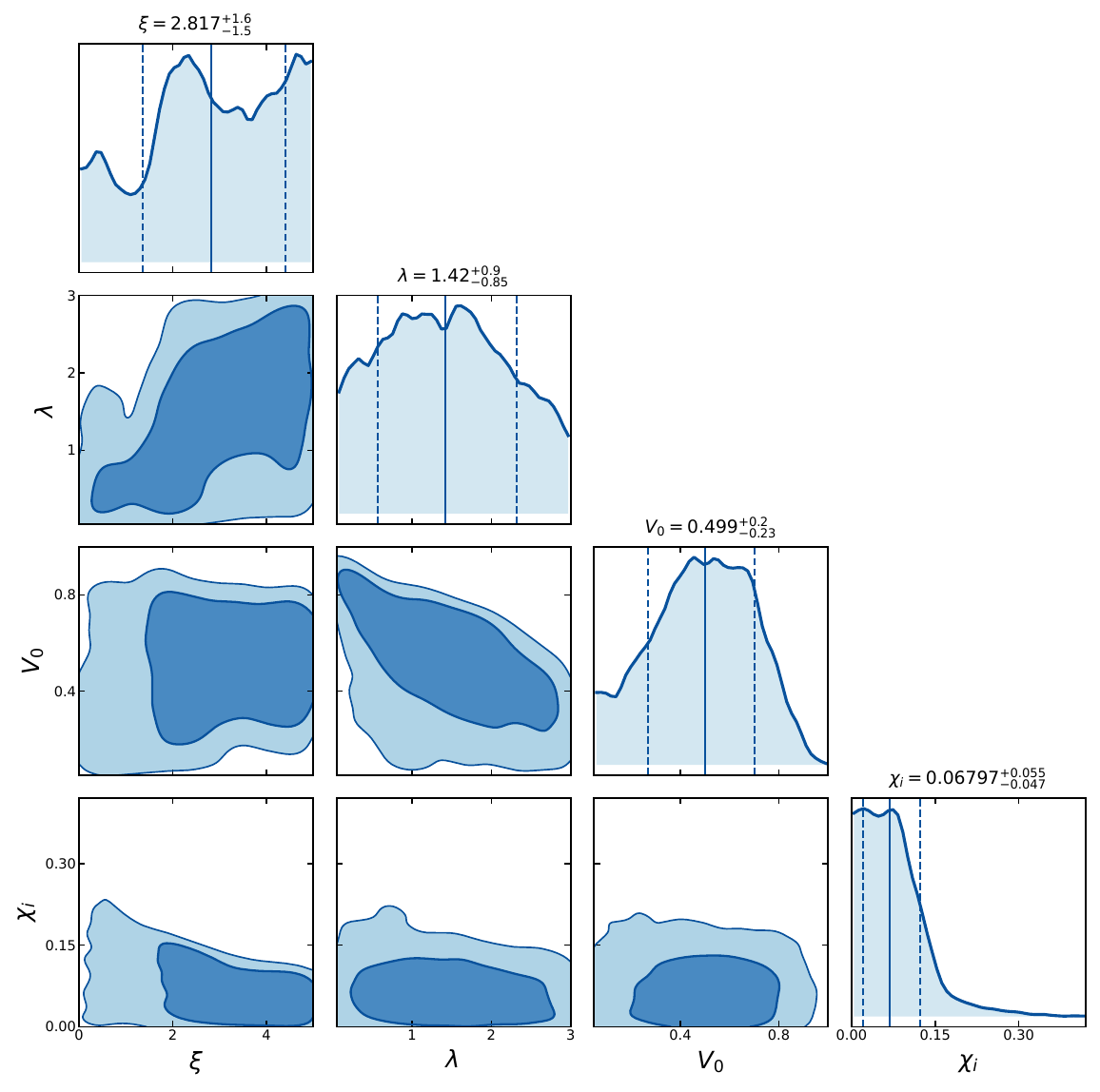}
\caption{Posterior distribution of the microscopic model parameters
$(\xi,\lambda,V_0,\chi_i)$ in the baseline prior range. The broad and curved
contours indicate degeneracies that remain after using background-level
reduced-CMB+BAO+SN data. This is not a plotting artifact: these parameters enter the
distance likelihood mainly through the derived background expansion history.
The intervals for these microscopic parameters should therefore be
interpreted as prior-conditional background-level constraints.}
\label{fig:model_posterior}
\end{figure*}

Figures~\ref{fig:posterior_bands} and \ref{fig:background_posterior} show
how the model realizes late-time phantom-crossing behavior while remaining
consistent with the distance data used in the likelihood. The posterior
median has $w_0>-1$ but evolves toward a mildly phantom regime at
intermediate redshift, whereas the expansion-rate band $E(z)$ is tightly
constrained. Together with the BAO and SN chi-square scales quoted
above, this demonstrates that the non-minimal coupling can generate the
desired late-time behavior without spoiling the reduced-CMB+BAO+SN
background constraints.

In terms of the inferred Hubble parameter, the combined distance data give
$H_0=67.88^{+0.53}_{-0.61}~{\rm km\,s^{-1}\,Mpc^{-1}}$. The analysis therefore
addresses the internal consistency of the phantom-crossing running-curvaton
background with CMB+BAO+SN distances, rather than the local-distance
$H_0$ discrepancy. The possibility that non-minimal coupling can improve
agreement with DESI-like late-time dynamics has been studied in recent work
\cite{Ye:2024ywg,Ye:2024thawing,Wolf:2025jed}; the role of the present
analysis is to embed this mechanism into the running-curvaton framework and
test the resulting homogeneous expansion history against the reduced
CMB+BAO+SN distance data.

\subsection{Convergence diagnostics and parameter degeneracy}

The convergence quantities are used here as descriptive diagnostics rather than as analytical results. For the 30000-step chain, the acceptance fraction ranges from 0.143 to 0.216, with median 0.190. The integrated
autocorrelation times are approximately
$\tau=(501,435,544)$ for $(h,\omega_b,\omega_c)$ and
$\tau=(714,604,615,696)$ for $(\xi,\lambda,V_0,\chi_i)$. The corresponding
$N_{\rm step}/\tau$ values are $(59.9,69.0,55.1)$ for
$(h,\omega_b,\omega_c)$ and $(42.0,49.7,48.8,43.1)$ for
$(\xi,\lambda,V_0,\chi_i)$. The trace of the log posterior reaches a
plateau, and the samples are not concentrated at the prior edges.
Nevertheless, because several microscopic parameters still have
$N_{\rm step}/\tau<50$, the trace plots and posterior summaries are stable
enough for a background-level consistency test, but not all parameters
satisfy the stricter $N_{\rm step}/\tau>50$ criterion normally desired for
final precision intervals.

The posterior-information summary gives the same message quantitatively. The
background and early-density parameters are strongly compressed by the
likelihood, while $\xi$, $\lambda$, $V_0$ and $\chi_i$ remain much broader
and show skewed, long-tailed marginalized contours in
Fig.~\ref{fig:model_posterior}. Here the prior-compression ratio is defined
as
\begin{equation}
{\cal R}_{68}
=\frac{q_{84}-q_{16}}{\theta_{\rm max}-\theta_{\rm min}},
\end{equation}
where $q_{16}$ and $q_{84}$ are the posterior percentiles and
$\theta_{\rm min}$, $\theta_{\rm max}$ are the top-hat prior bounds. This
ratio measures how much the likelihood compresses a parameter relative to its
allowed prior interval; a smaller value indicates a more informative
constraint. For $\xi$, the 68\% posterior width is
$1.468+1.582=3.05$, while the prior width is 5, giving
${\cal R}_{68}\simeq0.61$. By contrast, the same calculation gives
${\cal R}_{68}\simeq0.028$ for $h$ and ${\cal R}_{68}\simeq0.020$ for
$\omega_c$. Thus the background data sharply compress $h$ and $\omega_c$,
while $\xi$ remains much closer to its prior width.

A supplementary check of the upper prior bound on $\xi$ is given in
Appendix~\ref{app:xi_prior_sensitivity}. This check is not part of the main
production posterior. It shows that widening the prior from $\xi<5$ to
$\xi<8$ broadens the high-$\xi$ tail, while the density in the extended
interval $5<\xi<8$ remains lower than in the main baseline-prior region.

\section{Discussion and Conclusion}
\label{discussion}

We have investigated whether a single non-minimally coupled running-curvaton
field can consistently connect the early curvaton mechanism with a viable
late-time dark-energy background. The non-minimal coupling shifts the
inflationary curvaton mass, but this shift can be absorbed into the effective
running-curvaton parameter at leading order. This retuning is the consistency
condition that allows us to keep the original curvaton perturbation picture
as the starting point while using the same field for late-time dynamics.

At late times, the geometric terms induced by $\xi\chi^2R$ modify the
effective dark-energy pressure and allow the background equation of state to
cross the phantom divide without introducing a phantom kinetic term. The
posterior band in Fig.~\ref{fig:posterior_bands} shows that such mildly
phantom histories are present in the region selected by the reduced
CMB+DESI DR2 BAO+Pantheon+ likelihood. At the same time, the tightly
constrained expansion-rate band and the posterior constraints in
Fig.~\ref{fig:background_posterior} show that these histories do not spoil
the distance information used in the fit. In this sense, the model passes a
background-level consistency test: it can realize DESI-motivated
phantom-crossing dark energy while remaining compatible with the compressed
CMB acoustic scale, BAO distances and Pantheon+ supernova distances.

In terms of the Hubble parameter, the combined distance data prefer
$H_0\simeq 67.9~{\rm km\,s^{-1}\,Mpc^{-1}}$, close to the CMB-inferred
range. The derived background
quantities are constrained much more tightly than the microscopic parameters
$(\xi,\lambda,V_0,\chi_i)$, whose marginalized contours in
Fig.~\ref{fig:model_posterior} remain broad and curved. This indicates a
genuine degeneracy of the background-level likelihood rather than a failure
of the phantom-crossing mechanism. The supplementary widened-prior check in
Fig.~\ref{fig:xi_prior_robustness} shows that the upper tail of $\xi$ is
prior sensitive, although the widened chain does not run away toward
$\xi=8$. The microscopic/NMC intervals should therefore be regarded as
prior-conditional background-level constraints.

Several limitations remain. The CMB information used here is compressed to
background-level quantities. It does not include the effect of the
non-minimal coupling on TT/TE/EE spectra, CMB lensing, the late integrated
Sachs-Wolfe effect, or scalar perturbation evolution. The local-gravity
discussion is also qualitative and must be replaced by a dedicated
Solar-System or screening analysis. A full Boltzmann-code implementation,
followed by a complete CMB+BAO+SN parameter estimation, is therefore the
necessary next step for turning the present consistency test into a
definitive constraint on the non-minimally coupled running-curvaton model.

\section*{Data and code availability}

The DESI DR2 BAO measurements and covariance matrix used in this work are
publicly available and were read from the Cobaya BAO data package
\cite{Torrado:2020dgo}, which is used as a packaging source for the public
DESI data and not as a separate observational project. The specific files are
{\tt desi\_gaussian\_bao\_ALL\_GCcomb\_mean.txt} and
{\tt desi\_gaussian\_bao\_ALL\_GCcomb\_cov.txt}. The supernova likelihood
uses the public Pantheon+ data and covariance files
\cite{Scolnic:2021amr,Brout:2022vxf}
{\tt Pantheon+SH0ES.dat} and {\tt Pantheon+SH0ES\_STAT+SYS.cov}, with
Cepheid calibrator rows removed in the baseline analysis. The scripts used
for the reduced-CMB+BAO+SN likelihood, MCMC diagnostics and figure generation
can be made available upon reasonable request.

\section{ACKNOWLEDGMENTS}
B.L. and L.-H.L. are supported by the National Natural Science Foundation of China (Grant No. 12665009) and Education Department of Hunan Province (NO. 25B0480).


\appendix

\section{Supplementary diagnostics for the reduced-CMB consistency check}
\label{app:reduced_cmb_diagnostics}

This appendix collects the detailed early-background diagnostics used to document the assumptions behind the reduced-CMB acoustic-scale prior. These quantities are auxiliary diagnostics only. They are not additional likelihood terms, and they should not be interpreted as a full recombination-era or perturbation-level CMB test.

The quantities involving $\chi(z_\ast)$, $F(z_\ast)/M_P^2$ and
$\Omega_\chi(z_\ast)$ are unavailable because the current late-time solve starts at $z_{\rm ini}=50$. This is a limitation of the background solver range used in the reduced-CMB consistency test, not a lack of observational data. A direct test of these quantities would require extending the scalar background and perturbation evolution through recombination.

\begin{widetext}
\begin{center}
\begingroup
\footnotesize
\setlength{\tabcolsep}{4pt}
\renewcommand{\arraystretch}{1.02}
\refstepcounter{table}
\label{tab:early_background_diagnostics}
\noindent\parbox{0.98\textwidth}{\textbf{TABLE~\thetable.}
Early-background consistency diagnostics for the 30000-step
reduced-CMB+DESI DR2 BAO+Pantheon+ chain. The quantities $z_\ast$, $z_d$
and $r_s(z_\ast)$ use the same fitting-formula and high-redshift
standard-background prescription as the reduced-CMB likelihood. Quantities
involving $\chi(z_\ast)$ are unavailable because the current late-time
solver starts at $z_{\rm ini}=50$ and assumes the curvaton/NMC contribution
is negligible at earlier times.}
\vspace{0.35em}

\newcommand{\diagnosticnote}[1]{\parbox[t]{3.05in}{\raggedright #1}}
\begin{tabular}{@{}lccc@{}}
\toprule
Quantity & Best fit & Posterior median & Note\\
\midrule
$z_\ast$ & 1092.006 & 1091.916 & \diagnosticnote{Hu-Sugiyama recombination fit}\\
$z_d$ & 1020.306 & 1020.377 & \diagnosticnote{Eisenstein-Hu drag-redshift fit}\\
$r_s(z_\ast)$ [Mpc] & 144.494 & 144.563 & \diagnosticnote{standard high-z integral}\\
$r_d$ [Mpc] & 147.462 & 147.505 & \diagnosticnote{DESI-style fitting formula}\\
$\theta_\ast$ & 0.0104102 & 0.0104107 & \diagnosticnote{reduced CMB prior entry}\\
$\chi^2_{\rm CMB}$ & 0.0652 & 2.009 & \diagnosticnote{compressed-prior contribution}\\
CMB pull: $\theta_\ast$ & 0.0645 & 0.2576 & \diagnosticnote{diagonal sigma}\\
CMB pull: $\omega_b$ & 0.0054252 & 0.3050 & \diagnosticnote{diagonal sigma}\\
CMB pull: $\omega_{bc}$ & -0.2198 & -0.5096 & \diagnosticnote{diagonal sigma}\\
$F(0)/M_P^2$ & 0.8150 & 0.9511 & \diagnosticnote{solved for $0<z<50$; 80 posterior samples}\\
$\min F/M_P^2$ & 0.8150 & 0.9492 & \diagnosticnote{minimum over $0<z<50$; 80 posterior samples}\\
$\max |F/M_P^2-1|$ & 0.1850 & 0.0508 & \diagnosticnote{over $0<z<50$; 80 posterior samples}\\
$\alpha_M(0)$ & -0.5505 & -0.1007 & \diagnosticnote{$d\ln F/d\ln a$ at $z=0$; 80 posterior samples}\\
$\Omega_\chi(z_{\rm ini}=50)$ & 0.0029584 & 0.0083892 & \diagnosticnote{effective estimate; 80 posterior samples}\\
$\Omega_\chi(z_\ast)$ & \multicolumn{1}{c}{--} & \multicolumn{1}{c}{--} & \diagnosticnote{unavailable in late-time solve}\\
$F(z_\ast)/M_P^2$ & \multicolumn{1}{c}{--} & \multicolumn{1}{c}{--} & \diagnosticnote{unavailable in late-time solve}\\
\bottomrule
\end{tabular}
\par
\endgroup
\end{center}
\end{widetext}

\section{Supplementary sensitivity to the upper prior on \texorpdfstring{$\xi$}{xi}}
\label{app:xi_prior_sensitivity}

The baseline analysis uses $0<\xi<5$. Since Fig.~\ref{fig:model_posterior}
shows a broad $\xi$ posterior, we made a supplementary check with only the upper bound widened to $\xi<8$. We then continued the chain for 5000 steps from the final baseline samples. This widened-prior run is only an exploratory prior-sensitivity check, not an independent production chain with the same statistical status as the baseline 30000-step run. It tests whether the baseline posterior is mainly set by the chosen upper bound.

\begin{figure}[b]
\centering
\includegraphics[width=0.95\columnwidth]{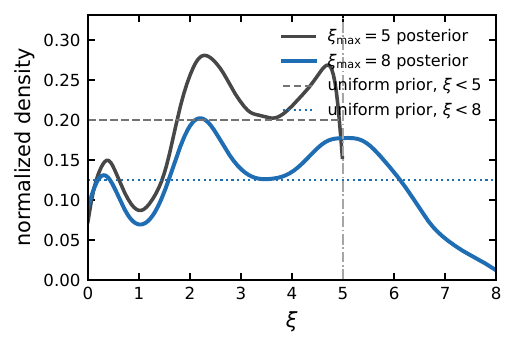}
\caption{Supplementary sensitivity check for the upper prior bound on the non-minimal coupling parameter $\xi$. The baseline posterior with $\xi_{\rm max}=5$ is compared with a short continuation using $\xi_{\rm max}=8$. The high-$\xi$ tail broadens when the prior is widened, but the density in the extended interval remains lower than in the main $0<\xi<5$ region. The widened-prior chain is used only to diagnose prior sensitivity of the high-$\xi$ tail.}
\label{fig:xi_prior_robustness}
\end{figure}

\newpage

The comparison is shown in Fig.~\ref{fig:xi_prior_robustness}. The median shifts from $\xi=2.829$ to $\xi=3.611$, corresponding to
$|\Delta\xi|/\sigma_{\rm wide}=0.37$. The wider prior allows a high-$\xi$ tail, but the posterior density in the extended interval $5<\xi<8$ is lower than in the main $0<\xi<5$ region. Only 3.2\% of the widened-prior samples lie above $\xi=7$. After accounting for the wider prior interval, the relative density also decreases toward the upper edge. We therefore use the baseline $\xi<5$ range as the reference prior for the baseline fit. The important point is not that $\xi$ is sharply measured, but rather that the distance likelihood weakly constrains this microscopic parameter while tightly
constraining the derived background expansion.

\section{Derivation of the Spectral Index with Non-minimal Coupling}
\label{app:spectral_index}

In this appendix, we provide a detailed derivation of the spectral index $n_{\chi}$ for the curvaton field non-minimally coupled to gravity. We show that the geometric correction to the effective mass leads directly to the modified spectral tilt used in our re-tuning mechanism.

\subsection{Perturbation Equation of Motion}
Starting from the action in the Jordan frame as given in \eqref{eq:action}, we consider the linear perturbations of the curvaton field $\chi(t, \mathbf{x}) = \bar{\chi}(t) + \delta\chi(t, \mathbf{x})$. In the flat FLRW background $ds^2 = -dt^2 + a^2(t)d\mathbf{x}^2$, the equation of motion for the Fourier mode $\delta\chi_k$ is given by:
\begin{equation}
    \ddot{\delta\chi}_k + 3H\dot{\delta\chi}_k + \left( \frac{k^2}{a^2} + m_{\rm eff}^2 \right) \delta\chi_k = 0,
    \label{eq:pert_eom}
\end{equation}
where the effective mass squared $m_{\rm eff}^2$ includes the geometric contribution from the non-minimal coupling:
\begin{equation}
    m_{\rm eff}^2 \equiv V''(\bar{\chi}) + \xi R.
\end{equation}
During slow-roll inflation, the Ricci scalar is dominated by the Hubble expansion, $R = 6(2H^2 + \dot{H}) \approx 12H^2$. Thus, the effective mass becomes:
\begin{equation}
    m_{\rm eff}^2 \approx V''(\bar{\chi}) + 12\xi H^2.
\end{equation}

\subsection{Reduction to Bessel Equation}
To solve Eq.~(\ref{eq:pert_eom}), we switch to conformal time $\tau = \int dt/a(t)$ and define the canonical variable $u_k(\tau) = a(\tau)\delta\chi_k(\tau)$. The evolution equation transforms into:
\begin{equation}
    u_k'' + \left( k^2 - \frac{a''}{a} + a^2 m_{\rm eff}^2 \right) u_k = 0.
\end{equation}
In the de Sitter limit ($a \approx -1/H\tau$), the effective potential term becomes $(a''/a - a^2 m_{\rm eff}^2) \approx \frac{1}{\tau^2}(2 - m_{\rm eff}^2/H^2)$. Consequently, the mode equation takes the form of a Bessel equation:
\begin{equation}
    u_k'' + \left( k^2 - \frac{\nu^2 - 1/4}{\tau^2} \right) u_k = 0,
\end{equation}
where the order $\nu$ is defined by:
\begin{equation}
    \nu = \sqrt{\frac{9}{4} - \frac{m_{\rm eff}^2}{H^2}}.
\end{equation}

\subsection{Power Spectrum and Spectral Index}
The solution matching the Bunch-Davies vacuum at early times ($|k\tau| \gg 1$) is given by the Hankel function of the first kind:
\begin{equation}
    u_k(\tau) = \frac{\sqrt{\pi}}{2} \sqrt{-\tau} H_{\nu}^{(1)}(-k\tau).
\end{equation}
On super-horizon scales ($|k\tau| \ll 1$), the asymptotic behavior of the Hankel function yields the power spectrum $\mathcal{P}_{\delta\chi} \equiv \frac{k^3}{2\pi^2} |\delta\chi_k|^2$:
\begin{equation}
    \mathcal{P}_{\delta\chi}(k) \propto k^{3-2\nu}.
\end{equation}
The spectral index is defined as $n_{\chi} - 1 \equiv d \ln \mathcal{P}_{\delta\chi} / d \ln k = 3 - 2\nu$. For a light field ($m_{\rm eff}^2 \ll H^2$), we expand $\nu$ to first order:
\begin{equation}
    \nu \approx \frac{3}{2} \left( 1 - \frac{2}{3} \frac{m_{\rm eff}^2}{3H^2} \right) = \frac{3}{2} - \frac{m_{\rm eff}^2}{3H^2}.
\end{equation}
Substituting this back into the expression for $n_{\chi}$, we arrive at the standard result modified by the effective mass:
\begin{equation}
    n_{\chi} - 1 \approx \frac{2}{3} \frac{m_{\rm eff}^2}{H^2}.
    \label{eq:spectral_index_final}
\end{equation}
Substituting the explicit form of the mass $m_{\rm eff}^2 = m_{\rm orig}^2 + 12\xi H^2$, we obtain:
\begin{equation}
    n_{\chi} - 1 \approx \frac{2}{3} \frac{m_{\rm orig}^2}{H^2} + 8\xi.
    \label{spectral index of curvaton}
\end{equation}
This linear dependence on $\xi$ necessitates the parameter re-tuning strategy ($g_0 \to g_0 - 2\xi$) discussed in Section III to preserve consistency with Planck observations.

\section{Detailed Calculation of Non-Gaussianity \texorpdfstring{$f_{NL}$}{fNL}}
\label{sec:fnl_calculation}

This appendix derives the specific form of the non-linearity parameter
$f_{NL}$ in the sudden-decay approximation.

\subsection{Curvature Perturbation Expansion}
The curvature perturbation $\zeta$ on uniform density hypersurfaces is related to the energy density fluctuation $\delta\rho$ by:
\begin{equation}
    \zeta = -H \frac{\delta\rho}{\dot{\rho}} = \frac{1}{3} \frac{\delta\rho}{\rho},
\end{equation}
assuming the universe is dominated by a fluid with equation of state $w \approx 0$ (oscillating curvaton) or considering the decay into radiation.

For the curvaton field $\chi$, the energy density is given by $\rho_\chi \approx m_{\rm eff}^2 \chi^2 / 2$. The density perturbation, expanded to second order in the field fluctuation $\delta\chi$, is:
\begin{align}
    \rho_\chi(\bar{\chi} + \delta\chi) &= \frac{1}{2}m_{\rm eff}^2 (\bar{\chi} + \delta\chi)^2 \nonumber \\
    &= \frac{1}{2}m_{\rm eff}^2 \bar{\chi}^2 \left( 1 + 2\frac{\delta\chi}{\bar{\chi}} + \left(\frac{\delta\chi}{\bar{\chi}}\right)^2 \right).
\end{align}
Thus, the density contrast is:
\begin{equation}
    \frac{\delta\rho_\chi}{\rho_\chi} = 2\frac{\delta\chi}{\bar{\chi}} + \left(\frac{\delta\chi}{\bar{\chi}}\right)^2.
\end{equation}

\subsection{Transfer to Adiabatic Perturbation}
Using the sudden decay approximation, the total curvature perturbation $\zeta$ is a weighted sum of the inflaton and curvaton perturbations. Assuming the inflaton contribution is negligible, $\zeta$ is given by:
\begin{equation}
    \zeta \approx r_{dec} \zeta_\chi = \frac{r_{dec}}{3} \frac{\delta\rho_\chi}{\rho_\chi},
\end{equation}
where $r_{dec} \equiv \left. \frac{3\rho_\chi}{3\rho_\chi + 4\rho_\gamma} \right|_{decay} \approx \frac{\rho_\chi}{\rho_{tot}}$ is the energy fraction of the curvaton at the time of decay (in the limit $r_{dec} \ll 1$).

Substituting the expansion of $\delta\rho_\chi/\rho_\chi$:
\begin{equation}
    \zeta = \frac{r_{dec}}{3} \left[ 2\frac{\delta\chi}{\bar{\chi}} + \left(\frac{\delta\chi}{\bar{\chi}}\right)^2 \right].
    \label{eq:zeta_expansion}
\end{equation}

\subsection{Identifying \texorpdfstring{$f_{NL}$}{fNL}}
We decompose $\zeta$ into a Gaussian part $\zeta_g$ (linear term) and a non-Gaussian correction. From Eq.~(\ref{eq:zeta_expansion}), the linear Gaussian part is:
\begin{equation}
    \zeta_g = \frac{2r_{dec}}{3} \frac{\delta\chi}{\bar{\chi}}.
\end{equation}
We can express the quadratic term $(\delta\chi/\bar{\chi})^2$ in terms of $\zeta_g$:
\begin{equation}
    \left(\frac{\delta\chi}{\bar{\chi}}\right)^2 = \left( \frac{3}{2r_{dec}} \zeta_g \right)^2 = \frac{9}{4r_{dec}^2} \zeta_g^2.
\end{equation}
Now, substituting this back into Eq.~(\ref{eq:zeta_expansion}):
\begin{equation}
    \zeta = \zeta_g + \frac{r_{dec}}{3} \left( \frac{9}{4r_{dec}^2} \zeta_g^2 \right) = \zeta_g + \frac{3}{4r_{dec}} \zeta_g^2.
\end{equation}

The standard definition of the local non-linearity parameter $f_{NL}$ is given by the expansion:
\begin{equation}
    \zeta = \zeta_g + \frac{3}{5} f_{NL} \zeta_g^2.
\end{equation}
Comparing the coefficients of the $\zeta_g^2$ term:
\begin{equation}
    \frac{3}{5} f_{NL} = \frac{3}{4r_{dec}} \quad \Longrightarrow \quad f_{NL} = \frac{5}{4r_{dec}}.
\end{equation}

\section{Tensor Perturbations with Non-minimal Coupling}
\label{tensor modes}

In this appendix, we verify the stability of the tensor perturbations and estimate the corrections to the primordial gravitational waves induced by the non-minimal coupling term $\xi \chi^2 R$.

\subsection{ Second-Order Action for Tensor Modes}

We consider the tensor perturbations to the metric in the Jordan frame, defined as $g_{ij} = a^2(\tau) (\delta_{ij} + h_{ij})$, where $h_{ij}$ satisfies the transverse-traceless conditions ($\partial^i h_{ij} = 0$, $h^i_i = 0$). The second-order action for tensor modes arising from the gravitational sector of Eq.~(1) is given by:
\begin{equation}
    S_{h} = \frac{1}{8} \int d\tau d^3x \, a^2 \Omega(\chi) \left[ (h_{ij}')^2 - (\nabla h_{ij})^2 \right],
    \label{eq:tensor_action}
\end{equation}
where primes denote derivatives with respect to conformal time $\tau$. The non-minimal coupling modifies the effective Planck mass squared:
\begin{equation}
    M_{\rm eff}^2(\chi) \equiv \Omega(\chi) = M_{P}^2 - \xi \chi^2.
\end{equation}

\subsection{Evolution Equation}

To diagonalize the kinetic term, we define the canonical variable $v_k(\tau) = z_T h_k(\tau)$ for each polarization mode, with the pump field defined as:
\begin{equation}
    z_T(\tau) = \frac{a(\tau)}{2} \sqrt{\Omega(\chi)}.
\end{equation}
The equation of motion (Mukhanov-Sasaki equation) for the mode function $v_k$ in Fourier space is:
\begin{equation}
    v_k'' + \left( k^2 - \frac{z_T''}{z_T} \right) v_k = 0.
\end{equation}
During slow-roll inflation, the curvaton field $\chi$ is effectively frozen (or evolving very slowly compared to the expansion), so $\Omega(\chi) \approx \text{const}$. In the de Sitter limit where $a(\tau) \approx -1/(H\tau)$, the effective mass term simplifies to:
\begin{equation}
    \frac{z_T''}{z_T} \approx \frac{a''}{a} \approx \frac{2}{\tau^2}.
\end{equation}
This indicates that the evolution of tensor modes remains effectively the same as in General Relativity, up to a rescaling of the normalization factor $z_T$.

\subsection{ Primordial Power Spectrum}

Solving the mode equation with Bunch-Davies vacuum initial conditions leads to the primordial tensor power spectrum on super-horizon scales:
\begin{equation}
    \mathcal{P}_{T}(k) = \frac{2 H^2}{\pi^2 M_{\rm eff}^2} = \frac{2 H^2}{\pi^2 (M_{P}^2 - \xi \chi_*^2)},
\end{equation}
where quantities are evaluated at the horizon exit. We can express this relative to the standard GR prediction $\mathcal{P}_{T, GR} \approx \frac{2 H^2}{\pi^2 M_{P}^2}$:
\begin{equation}
\begin{aligned}
\mathcal{P}_{T}(k)
&= \mathcal{P}_{T,{\rm GR}}
\left[1-\xi\left(\frac{\chi_*}{M_P}\right)^2\right]^{-1}\\
&\simeq
\mathcal{P}_{T,{\rm GR}}
\left[1+\xi\frac{\chi_*^2}{M_P^2}\right].
\end{aligned}
    \label{tensor mode}
\end{equation}

\subsection{ Consistency Check: Sub-dominance Condition}

The validity of our model relies on the assumption that the curvaton $\chi$ is a spectator field during inflation, meaning its energy density is negligible compared to the inflaton, and its field value is sub-Planckian:
\begin{equation}
    \chi_* \ll M_P \quad \Longrightarrow \quad \frac{\chi_*^2}{M_P^2} \ll 1.
\end{equation}
Assuming the field does not traverse trans-Planckian distances from inflation to today, the correction term $\xi (\chi/M_P)^2$ remains negligible throughout cosmic history.

\section{Derivation of the Effective Equation of State}
\label{derivation of effective equation of state}
We consider the running-curvaton model extended with a non-minimal
gravitational coupling in the Jordan frame. The action is given by
\eqref{eq:action},
where the non-minimal coupling function is defined as:
\begin{equation}
\Omega(\chi) = M_P^2 - \xi \chi^2.
\end{equation}
We vary the action with respect to the metric. For the gravitational sector,
$\delta S_g=\frac{1}{2}\int d^4x\,\delta\left(\sqrt{-g}\Omega R\right)$,
one needs
\begin{align}
\delta(\sqrt{-g}\Omega R)
&=(\delta\sqrt{-g})\Omega R+\sqrt{-g}\Omega\,\delta R \nonumber\\
&=-\frac{1}{2}\sqrt{-g}g_{\mu\nu}\Omega R\,\delta g^{\mu\nu}
\nonumber\\
&\quad+\sqrt{-g}\Omega R_{\mu\nu}\delta g^{\mu\nu}
\nonumber\\
&\quad+\sqrt{-g}\Omega g_{\mu\nu}\Box\delta g^{\mu\nu}
\nonumber\\
&\quad-\sqrt{-g}\Omega\nabla_\mu\nabla_\nu\delta g^{\mu\nu}.
\label{varying with gravitational}
\end{align}
Since $\Omega(\chi)$ is not a constant, the total divergence term in $\delta R$ cannot be discarded. We perform integration by parts to shift the derivatives onto $\Omega$:
\begin{align}
&\int d^4x \sqrt{-g}\,\Omega
(g_{\mu\nu}\Box-\nabla_\mu\nabla_\nu)\delta g^{\mu\nu}
\nonumber\\
&\quad =
\int d^4x \sqrt{-g}\,
(g_{\mu\nu}\Box\Omega-\nabla_\mu\nabla_\nu\Omega)
\delta g^{\mu\nu}.
\end{align}
Combining these terms, the variation of the gravitational sector is:
\begin{equation}
\begin{aligned}
\frac{\delta S_g}{\delta g^{\mu\nu}}
&= \sqrt{-g}\Big[
\frac{1}{2}\Omega
\left(R_{\mu\nu}-\frac{1}{2}g_{\mu\nu}R\right)\\
&\quad
+\frac{1}{2}
\left(g_{\mu\nu}\Box\Omega-\nabla_\mu\nabla_\nu\Omega\right)
\Big].
\end{aligned}
\end{equation}

Next, varying the canonical scalar part gives
\begin{equation}
T_{\mu\nu}^{(\chi)}
=\nabla_\mu\chi\nabla_\nu\chi
-g_{\mu\nu}
\left[\frac{1}{2}(\nabla\chi)^2+V(\chi)\right].
\end{equation}
Combining the above results gives the Jordan-frame modified Einstein equation,
\begin{equation}
\begin{aligned}
\Omega(\chi) G_{\mu\nu}
&= T_{\mu\nu}^{(m)} + T_{\mu\nu}^{(\chi)}
+ \nabla_\mu \nabla_\nu \Omega(\chi)\\
&\quad - g_{\mu\nu} \Box \Omega(\chi),
\end{aligned}
\label{modify_field_eq}
\end{equation}
where $T_{\mu\nu}^{(\chi)} = \nabla_\mu \chi \nabla_\nu \chi - g_{\mu\nu} [\frac{1}{2}(\nabla\chi)^2 + V(\chi)]$ is the energy-momentum tensor of the scalar field.

For the flat FLRW metric in Eq.~\eqref{frw metric}, the non-vanishing Einstein
tensor components are
\begin{equation}
\begin{aligned}
G_{00}&=3H^2,\\
G_{ij}&=-a^2\eta_{ij}(-3H^2-2\dot H),
\end{aligned}
    \label{einstein tensor}
\end{equation}
For a homogeneous field $\chi(t)$,
\begin{equation}
\begin{aligned}
\Box\Omega&=-\ddot{\Omega}-3H\dot{\Omega},\\
\nabla_0\nabla_0\Omega&=\ddot{\Omega},\\
\nabla_i\nabla_j\Omega&=-a^2H\delta_{ij}\dot{\Omega}.
\end{aligned}
\end{equation}

The $00$ component of Eq.~\eqref{modify_field_eq} is
\begin{equation}
3(M_P^2 - \xi \chi^2) H^2 = \rho_m + \frac{1}{2}\dot{\chi}^2 + V(\chi) + 6\xi H \chi \dot{\chi}.
\label{eq:Friedmann1_derived}
\end{equation}
The $ij$ component is
\begin{align}
\Omega(\chi)(-2\dot H-3H^2)
&=p_m+\frac{1}{2}\dot{\chi}^2 - V(\chi) \nonumber\\
&\quad
-2\xi(\dot{\chi}^2+\chi\ddot{\chi})
-4\xi H\chi\dot{\chi}.
    \label{ij component of einstein eq}
\end{align}
During these calculations, we have utilized 
\begin{align}
\dot{\Omega} &= -2\xi \chi \dot{\chi}, \\
\ddot{\Omega} &= -2\xi (\dot{\chi}^2 + \chi \ddot{\chi}).
\end{align}
We recast the modified Friedmann equations into the standard Einsteinian form to define effective dark energy quantities:
\begin{align}
3M_P^2 H^2 &= \rho_m + \rho_{DE}^{eff}, \\
-2M_P^2 \dot{H} &= (\rho_m + p_m) + (\rho_{DE}^{eff} + p_{DE}^{eff}).
\end{align}
These definitions lead to the effective dark-energy density
\eqref{energy density of de} and pressure \eqref{presure of de}.

\bibliographystyle{apsrev4-2}
\bibliography{references}

\end{document}